\documentclass[conference,10pt,a4paper]{IEEEtran}
\IEEEoverridecommandlockouts

\makeatletter
\def\markboth#1#2{\def\leftmark{\@IEEEcompsoconly{\sffamily}\MakeUppercase{\protect#1}}%
\def\rightmark{\@IEEEcompsoconly{\sffamily}\MakeUppercase{\protect#2}}}
\makeatother

 \PassOptionsToPackage{bookmarks={false}}{hyperref}
\usepackage[utf8x]{inputenc}
\usepackage[english]{babel}
\usepackage{ucs}
\usepackage{amsmath}
\usepackage{amsfonts}
\usepackage{amssymb}
\usepackage{amsthm}
\usepackage{array}
\usepackage{verbatim}
\usepackage{listings}
\usepackage{psfrag}
\usepackage{stfloats}

\usepackage{algorithm}
\usepackage{algorithmic}
\usepackage{url}
  \usepackage{enumerate}
  \usepackage{cite}
\usepackage[usenames,dvipsnames,svgnames,table]{xcolor}

\ifCLASSINFOpdf
  \usepackage[pdftex]{graphicx}
  \usepackage{subfigure}
  \usepackage{epstopdf}
   \graphicspath{{./img/}{../matlab/MFBPsim/figures/}}
   \DeclareGraphicsExtensions{.eps,.ps,.png}
\else
  \usepackage[dvipdf]{graphicx}
  \usepackage{subfigure}
  \graphicspath{{./img/}}
   \DeclareGraphicsExtensions{.eps,.ps}
\fi

\newcommand{\Hb}{\mathbf{H}}
\newcommand{\Sb}{\mathbf{\Sigma}}
\newcommand{\U}{\mathbf{U}}

\newcommand{\V}{\mathbf{V}}

\newcommand{\I}{\mathbf{I}}

\newcommand{\D}{\mathbf{D}}

\newcommand{\x}{\mathbf{x}}

\newcommand{\vv}{\mathbf{v}}
\newcommand{\uu}{\mathbf{u}}
\newcommand{\y}{\mathbf{y}}

\newcommand{\n}{\mathbf{n}}

\newcommand{\Ex}[2]{{\textnormal{E}_{#1}\left[#2\right]}}

  \usepackage{xcolor}
   \definecolor{blueH3}{rgb}{0,.5,1}
   \definecolor{blueH2}{rgb}{0,0.25,0.75}
   \definecolor{blueH1}{rgb}{0,0,0.5}
   \definecolor{grayOldText}{rgb}{.5,.5,.5}
   \definecolor{VCobalt}{HTML}{005682}
   \definecolor{TZTeal}{HTML}{008080}
   \definecolor{KYJade}{HTML}{008151}
   \definecolor{ARust}{HTML}{a10000}

\title{Bit Allocation for Increased Power Efficiency in 5G Receivers with Variable-Resolution ADCs}

\author{
    \IEEEauthorblockN{Waqas~bin~Abbas\IEEEauthorrefmark{1}, Felipe~Gomez-Cuba\IEEEauthorrefmark{2}, Michele Zorzi\IEEEauthorrefmark{1}
    \IEEEauthorblockA{
    \IEEEauthorrefmark{1}DEI, University of Padua, Italy.\\
   \texttt{\{waqas,zorzi\}@dei.unipd.it}\\
    \IEEEauthorrefmark{2}AtlantTIC, University of Vigo, Spain.\\
    \texttt{fgomez@gti.uvigo.es}
    }
    \vspace{-1cm}}
   
    \thanks{F. Gomez-Cuba's work has been partially supported by FPU2012/01319.}
    \thanks{Michele Zorzi's work has been partially supported by NYU-Wireless.}
    \thanks{This project has received funding from the European Union's Horizon 2020 research and innovation programme under the Marie Sk\l{}odowska-Curie grant agreement No 704837.}
}

\begin{document} 
\maketitle

\begin{abstract}
In future high-capacity wireless systems based on mmWave or massive multiple input multiple output (MIMO), the power consumption of receiver Analog to Digital Converters (ADC) is a concern. Although hybrid or analog systems with fewer ADCs have been proposed, fully digital receivers with many lower resolution ADCs (and lower power) may be a more versatile solution. In this paper, focusing on an uplink scenario, we propose to take the optimization of ADC resolution one step further by enabling variable resolutions in the ADCs that sample the signal received at each antenna. This allows to give more bits to the antennas that capture the strongest incoming signal and fewer bits to the antennas that capture little signal energy and mostly noise. Simulation results show that, depending on the unquantized link SNR, a power saving in the order of 20-80\% can be obtained by our variable resolution proposal in comparison with a reference fully digital receiver with a fixed low number of bits in all its ADCs.
\end{abstract}


\section{Introduction}
\label{sec:introduction}
Future wireless communications are expected to leverage large antenna arrays at the base station to achieve higher data rates, both in new  millimeter wave (mmWave) bands and at standard frequencies with massive multiple input multiple output (MIMO) \cite{KhanFmmWave,MassiveMIMO_Mag_Marzetta}. 
Fully digital receiver architectures, where each antenna is connected to an independent Analog to Digital Converter (ADC), can provide maximum flexibility but could display too high component power consumptions due to the exponential increase of ADC power with the number of bits \cite{mmWBF2014}. The concept of green communication and the deployment of ultra-dense small cells motivate the reduction of power consumption of the base station.

There are two strategies to mitigate the power consumption of receivers with many antennas:
\begin{enumerate}
 \item Use Analog or Hybrid Combining (AC or HC) to perform all or a part of the MIMO operations in analog circuitry and sample only one or a few signals with ADCs \cite{AlkhateebMIMOSolMag,AyachHC}.
 \item Use fully Digital Combining (DC) with reduced ADC resolution (for example, 1 or a few bits), which can offer even better power efficiency if the power of radio-frequency (RF) components is taken into account \cite{MoHeath1bit,ZhangSELowADC,MyPCCompEW16}.
\end{enumerate}

In this work, focusing on an uplink scenario, we propose a further improvement to the fully-digital low-resolution strategy by studying the possibility of enabling a variable number of bits in each ADC of the DC system. Compared to a conventional approach to low-resolution DC, where each RF chain has equal ADCs with the same fixed number of bits $b_{\mathrm{ref}}$, we propose assigning to some ADCs a slighly higher number of bits $b_{\mathrm{high}}>b_{\mathrm{ref}}$, while the rest of the RF chains have an even lower number of bits ($b_{\mathrm{low}}<b_{\mathrm{ref}}$). Our results show that the same capacity of the fixed-bit system can be achieved using two variable-bit values with a power saving between 20 and 80\%, depending on the link pre-quantization Signal to Noise Ratio (SNR).

\subsection{Related Work} 
\label{ssec:Rel_Work}

Recent works such as \cite{confiwcmcNossekI06,MurrayAGCQuant,AmineGaussQaun} study the capacity and energy efficiency (EE) of large antenna array receiver designs depending on the ADC resolution. The effect of the number of ADC bits $b$ and sampling rate $B$ on capacity and power consumption is analyzed in \cite{OrhanER15PowerCons} for both AC and DC.

DC systems using low-resolution ADCs to reduce power consumption are further analyzed in \cite{FanULRateLowADC15,ZhangSELowADC}, showing that a few bits are enough to achieve almost the spectral efficiency (SE) of an unquantized system of the same characteristics. 

It is possible to use analog switches instead of analog mixers to create a hybrid scheme that samples only the best subset of the antennas of the array to reduce power consumption, as proposed in \cite{RialandHeath}. In addition to switching the best antennas to high resolution ADCs, it is possible to add 1-bit ADCs to sample the rest of the antennas as in \cite{MixedADC}, achieving a large fraction of the capacity of a full-high-resolution architecture.

In this work we show that these configurations with antenna selection (equivalent to 0 bit ADCs) or only 1 bit, combined with a few very high resolution ADCs, are not necessarily optimal, and a milder variation in the number of bits such as $b_{\mathrm{low}}=4$ vs $b_{\mathrm{high}}=6$ may work better.

\subsection{Our Contribution}
\label{ssec:Our_Cont}
In this work, we focus on low resolution ADCs and discuss how the availability of variable resolution ADCs can reduce the receiver power consumption compared to a receiver with fixed resolution on each ADC.
We primarily focus on a simple case where the ADCs offer only two operation modes, with low and high resolutions. We believe such simplified model is a good starting point to open this topic and can be more easily imagined as a practically feasible hardware.

We propose two different algorithms depending on whether ADCs can only take two resolution values, or can also be completely shut off to also incorporate the benefits of antenna selection. We perform an analysis of power consumption under the constraint that the variable-resolution scheme achieves the same SNR after quantization and capacity compared to the reference fixed-resolution model. 
We show that
\begin{itemize}
\item A receiver with two-level ADC resolution can achieve similar capacity to a fixed resolution scheme such as \cite{FanULRateLowADC15,ZhangSELowADC} in practical systems with high SNR. However, instead of 2 bits like in \cite{FanULRateLowADC15} our reference is the fixed resolution that achieves the best trade-off between SE and EE under our channel model, which is  $b_{\mathrm{ref}}=5$ according to the results in \cite{WaqasEConJour16}\footnote{This may be visualized using the tool available at, http://enigma.det.uvigo.es/\textasciitilde{}fgomez/mmWaveADCwebviewer/, by selecting the HPADC (High-Power ADC) option and the values of the parameters discussed later in this paper.}.
\item The power saving is related to the two levels of resolution selected, $b_{\mathrm{low}}$ and $b_{\mathrm{high}}$. Robust performance at low SNR is obtained for not-too-low $b_{\mathrm{low}}$ and not-too-high $b_{\mathrm{high}}$, whereas at high SNR more power is saved with more extreme differences between $b_{\mathrm{low}}$ and $b_{\mathrm{high}}$.
\item The power saving increases slightly with a larger number of antennas, and so enabling variable resolution is even more interesting in massive MIMO systems.
\end{itemize} 

\section{System Model}
\label{sec:model}
\subsection{mmWave Channel}

We study mmWave point-to-point uplink MIMO links with an $N_t$ antenna transmitter, an $N_r$ antenna receiver and bandwidth $B$. We assume that there is no inter-symbol interference, as in previous models such as \cite{mustafa2013mmWave}. The received signal in each symbol period $1/B$ is

\begin{equation}
\textbf{y} = \textbf{H}\textbf{x} + \textbf{n}
\label{eq:y}
\end{equation}
where $\textbf{x}$ represents the transmitted symbol vector, $\textbf{n}$ is the independent and identically distributed (i.i.d) circularly symmetric complex Gaussian noise vector, $\textbf{n} \sim \mathcal{CN}(\textbf{0},N_o\textbf{I})$, where $N_o$ represents the noise power, and $\textbf{H}$ represents the $N_r \times N_t$ channel matrix. The mmWave channel
matrix $\textbf{H}$ is randomly distributed following a random geometry with a small number of propagation
paths (order of tens) grouped in very few clusters of similar paths \cite{mustafa2013mmWave}, and is obtained as

\begin{equation}
   \textbf{H} = \sqrt{\dfrac{N_{t}N_{r}}{\rho N_cN_p}}\sum_{k=1}^{N_c}\sum_{\ell=1}^{N_p}g_{k,\ell}\textbf{a}_{r}(\phi_{k}+\Delta\phi_{k,\ell}) \textbf{a}_{t}^H(\theta_k+\Delta\theta_{k,\ell})
\end{equation}
where the terms in this expression are generated according to the mmWave channel model in \cite{WaqasEConJour16}. Here $\rho$ is the pathloss, $g_{k,\ell}$ is the small scale fading coefficient associated with the $\ell^{th}$ path of the $k^{th}$ cluster,  $\textbf{a}_{t}$ and $\textbf{a}_{r}$ are spatial signatures of the transmit and receive arrays, and the $\theta$'s and $\Delta \theta$'s and $\phi$'s and $\Delta \phi$'s are the angles of departure and arrival for a small number of propagation paths $N_p$ grouped in even fewer independent clusters $N_c$.  

It must be noted that, due to this small number of paths, despite having large dimensions, the matrix $\Hb$ has a low rank and an even lower number of dominant eigenvalues are responsible for 95\% of the energy transfer in the channel. $\Hb$ is generated in \cite{mustafa2013mmWave} using $N_c\sim\mathrm{Poisson}(1.8)$ and $N_p=20$. In this paper, we generate $\Hb$ instead with $N_c=2$, $N_p=10$ for absolute compatibility with fixed-resolution power consumption values obtained in \cite{WaqasEConJour16}, where $N_c$ is selected as a constant and varied to study its effect. Moreover, it is noted in \cite{mustafa2013mmWave} that for the median channel a single spatial dimension captures approximately 50\% of the channel energy and two degrees of freedom capture 80\% of the channel energy. We also performed our own Monte-Carlo verification with $10^4$ channel realizations, and found that the first eigenvalue is responsible for over 50\% energy transfer with probability $0.95$ and for over 75\% of the energy transfer with probability $0.6$. 

Therefore, for the sake of space and simplicity, in our analysis we assume that the transmitter has the channel state information (CSIT) and implements a beamforming scheme that concentrates all the signal in the single strongest eigenvalue of the channel matrix. That is, if $\Hb=\U\Sb\V^H$ is the Singular Value Decomposition of the channel, the transmitter sends a scalar symbol, $x$, projected over the row $\vv_m$ of $\V^H$ associated with the strongest eigenvalue on the diagonal $\Sb$. Thus the signal at the transmitter array is $\x=\vv_m x$ and the received signal may be expressed as

\begin{equation}
 \y=\uu_m \sigma_m x +  \textbf{n}
\end{equation}

where $\sigma_m$ is the maximum singular value, and $\bf{u}_m$ is the corresponding left singular vector.

\begin{figure}[t]
 \centering
 \includegraphics[width=.75\columnwidth]{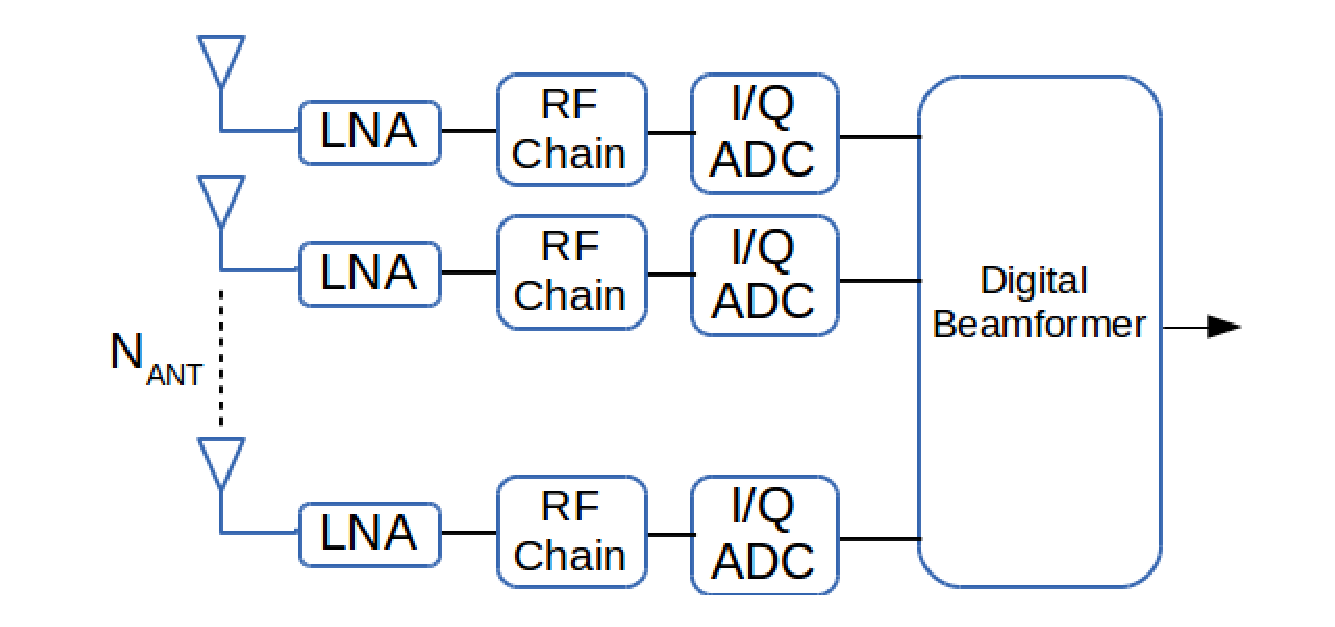}
 \caption{Digital Receiver with ADC with $b_i$ bits on each antenna RF chain.}
 \label{fig:receiver}
\end{figure}

\subsection{Variable-bits ADC Receiver}

\begin{figure}[t]
 \centering
 \includegraphics[width=0.45\columnwidth]{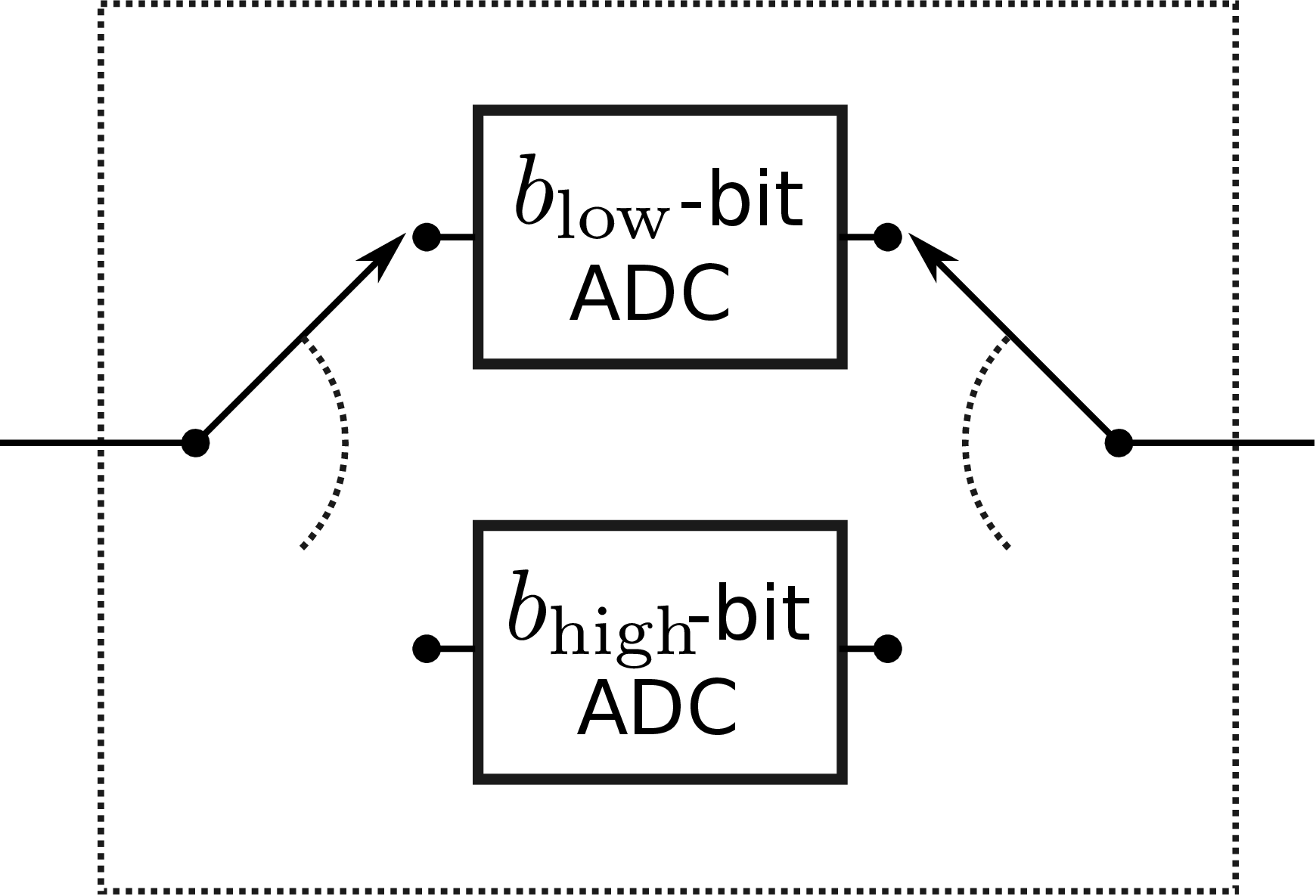}
 \caption{A simplified 2-level variable-resolution ADC.}
 \label{fig:varBADC}
\end{figure}

The DC receiver is illustrated in Fig. \ref{fig:receiver}. After the signal \eqref{eq:y} is received, the signal at each antenna $i$ is quantized by an ADC with $b_i$ bits. Due to the fact that we are only concerned about power consumption in this paper, but not the maximum number of components, we illustrate the variable resolution ADC with the simplified architecture in Fig. \ref{fig:varBADC}, consisting in a pair of fixed-resolution ADCs that can be alternatively switched in and out of the circuit. This simplified architecture serves to illustrate some interesting gains and open the discussion on variable resolution ADCs, while we leave the design of resource-efficient variable-resolution ADC architectures for future research. Moreover, the use of switching hardware guarantees that ADCs can be commuted with the same time resolutions used in antenna selection schemes. 

We represent the signal after quantization using the Additive Quantization Noise Model (AQNM) \cite{OrhanER15PowerCons} approximation by adding an additive white noise $\textbf{n}_q$ that models the quantization distortion of each coefficient $y_i$ of the signal \eqref{eq:y}, producing a quantized output in each ADC that satisfies
\begin{equation}
y_i^q = (1-\eta_i)y_i + n^q_i
\label{eq:yqi}
\end{equation}
where $\eta_i$ is the inverse of the signal-to-quantization noise ratio at antenna $i$, and is inversely proportional to the square of the resolution of the $i$-th ADC (i.e., $\eta_i \propto 2^{-2b_i}$). The quantization noise in each antenna $n^q_i$ is AWGN distributed with variance $\eta_i(1-\eta_i)\Ex{}{|y_i|^2}$.

We can write the quantized signal as a vector by denoting the quantization terms in a diagonal matrix, producing
\begin{equation}
\begin{split}
\y^q&=\D(\Hb\x+\n)+\n^q,\\
&\D=\left(\begin{array}{cccc}
				      (1-\eta_1)&0\dots&0\\
				      0&(1-\eta_2)\dots&0\\
				      \vdots&\vdots&\ddots&\vdots\\
				      0&0&\dots&(1-\eta_{N_r})\\
                                 \end{array}
\right),\\
\end{split}
\end{equation}
where $\D=(1-\eta_{\mathrm{ref}})\I$ when the number of bits is the same in all ADCs.

For a Gaussian input distribution, the values of $\eta$ for $b \leq 5$ are listed in Table \ref{tab:etavsb}, and for $b > 5$ can be approximated by  $\eta = \frac{\pi \sqrt{3}}{2} 2^{-2b}$ \cite{FanULRateLowADC15}. 
We denote by $\gamma_q$ the SNR of $\textbf{y}_q$, given by
 \begin{align}
	\gamma_q &= \bigg|\dfrac{(\D\textbf{H}\textbf{R}_{\textbf{xx}}\textbf{H}^H\D^H)}{\D^2{N_o} + \textbf{R}_{\textbf{n}^q\textbf{n}^q}}\bigg| \\
	&= \bigg|\dfrac{(\D\textbf{H}\textbf{R}_{\textbf{xx}}\textbf{H}^H\D^H)}{\textbf{D}^2{N_o} + ((\I-\D)(\textbf{H}\textbf{R}_{\textbf{xx}}\textbf{H}^H + N_0\textbf{I})\D^H)}\bigg|
	\label{eq:gammaQ}
\end{align}

  \begin{table}
     \centering
     \caption{$\eta$ for different values of $b$ \cite{FanULRateLowADC15}}
     \begin{tabular}{|c||c|c|c|c|c|}
         \hline
         $b$ &  1 & 2 & 3 & 4 & 5\\ 
         \hline
         $\eta$  & 0.3634 & 0.1175 & 0.03454 & 0.009497 & 0.002499 \\
         \hline
     \end{tabular}
     \label{tab:etavsb}\vspace{-3mm}
 \end{table}

where $\bf{R}_{n{_q}n{_q}}$ and $\textbf{R}_{\textbf{xx}}$ represent the covariance matrices of the quantization noise and of the transmitted symbol, respectively. Note that \eqref{eq:gammaQ} is valid for any type of transmission, and in this paper we can simplify it for our particular choice of single dominant eigenvalue beamforming. We replace $\textbf{R}_{\textbf{xx}}=|x|^2\vv_m\vv_m^H$, and the optimal receiver for this transmission is Maximum Ratio Combining so the SNR becomes
 \begin{equation}
 \begin{split}
	\gamma_q 
	&=\sum_i\frac{\sigma^2_m  |u_m^i|^2(1-\eta_i)^2}{ N_0(1-\eta_i)^2 + (N_0 +\sigma^2_m|u_m^i|^2)\eta_i(1-\eta_i)}
	\label{eq:gammaQ1}
\end{split}
\end{equation}

By imposing the constraint of using single dominant eigenvalue beamforming, we are able to express the SNR as a sum of per-RF-chain partial SNR's as $\gamma_q=\sum_i \gamma_q^i$, where the quantized signal from each antenna has partial SNR $\gamma_q^i=\frac{\sigma^2_m  |u_m^i|^2(1-\eta_i)^2}{ N_0(1-\eta_i)^2 + (N_0+\sigma^2_m|u_m^i|^2)\eta_i(1-\eta_i)}$. Also we can define the unquantized SNR per antenna $\gamma_i=\frac{|u_m^i|^2 \sigma^2_m}{N_0}$ such that $\gamma_q^i=\frac{(1-\eta_i)\gamma_i}{1+\eta_i\gamma_i}$.

Finally, the capacity of the MIMO link, which in general is

 \begin{equation}
	C_q = \mathrm{E}_{\mathbf{H}}\left[\max_{\textbf{R}_{\mathbf{xx}}}B\log_2\bigg|\textbf{I} + \dfrac{(1-\eta)(\textbf{H}\textbf{R}_{\textbf{xx}}\textbf{H}^H)}{{N_o}\textbf{I} + \eta(\textbf{H}\textbf{R}_{\textbf{xx}}\textbf{H}^H)}\bigg|\right],
	\label{eq:Cq_nor}
\end{equation}
can be particularized with our transmitter constraints to
 \begin{equation}
C_q =\Ex{\Hb}{\log (1+\gamma_q)}
\end{equation}

where $\Ex{}{.}$ represents the expectation. 

\subsection{Digital Receiver Power Consumption}


The devices required to implement the mmW receiver architecture are displayed in Fig. \ref{fig:receiver}. All receiver schemes considered in this paper have the same RF components and we are only interested in the variation of ADC power consumption as a function of their number of bits. 

The power consumption of the $i$-th ADC, denoted as $P_{ADC}^{i}=cB2^{b_i}$, increases exponentially with the number of bits $b_i$ and linearly with the bandwidth $B$ and with the ADC Walden's figure of merit $c$ \cite{ADC_b_B} (the energy consumption per conversion step per Hz). The aggregate power consumption across all the ADCs in the system is
\begin{equation}
   P^{Tot}_{ADC} =  \sum_{i} P_{ADC}^{i}= cB\left(\sum_i 2^{b_i}\right)
   \label{eq:P_ADC}
\end{equation} 
where it must be noted that we consider that all ADCs have the same Walden's figure of merit despite the variation of bits.

\section{Analysis of the Variable Bit ADC System}

We compare a reference receiver with a fixed number of bits in all ADCs, $b_\mathrm{ref}$, and a receiver where the number of bits on each ADC can be selected at any time instant  between two possible values, $b_\mathrm{low}$ and $b_{\mathrm{high}}$, depending on the instantaneous unquantized SNR's in each of the antennas $\gamma_i$.

With regard to the capacity of the channel, we must note that as $b_i$ grows, $\eta_i$ decreases exponentially to $0$. Each term in \eqref{eq:gammaQ1} of the form $\gamma_q^i=\frac{\sigma^2_m  |u_m^i|^2(1-\eta_i)^2}{ N_0 + \sigma^2_m|u_m^i|^2\eta_i(1-\eta_i)}$ increases monotonically as $\eta_i\to0$. Thus, the per-antenna SNRs always increase with $b_i$. Moreover, the higher $|u_m^i|^2$, the more the increase in SNR derived from assigning more bits to the $i$-th antenna.

With regard to the power consumption, denoting the number of antennas with  $b_{\mathrm{high}}$ bits by $N_{\mathrm{high}}$, we write the normalized power consumption (ratio between the power consumption with variable and fixed resolution) as

\begin{equation}\label{eq:normpow}
\begin{split}
 \xi&=\frac{cB (N_{\mathrm{high}} 2^{b_{\mathrm{high}}}+(N_r-N_{\mathrm{high}})2^{b_{\mathrm{low}}})}{cB N_r 2^{b_{\mathrm{ref}}}}\\
    &=\frac{N_{\mathrm{high}}}{N_r}2^{b_{\mathrm{high}}-b_{\mathrm{ref}}}+\left(1-\frac{N_{\mathrm{high}}}{N_r}\right)2^{b_{\mathrm{low}}-b_{\mathrm{ref}}}
 \end{split}
\end{equation}

The design of our variable-bit ADC system must satisfy that
\begin{itemize}
 \item In order to be able to replicate the capacity and/or power consumption of the reference system, we select bit values $$b_{\mathrm{low}}\leq b_{\mathrm{ref}}\leq b_{\mathrm{high}}.$$
 \item If we wish for the variable-bit system to consume less, or equal power as the reference system, we must have 
 \begin{equation}
 N_{\mathrm{high}}\leq 
 \frac{2^{b_{\mathrm{ref}}-b_{\mathrm{low}}}-1}{2^{b_{\mathrm{high}}-b_{\mathrm{low}}}-1}N_r
 \label{eq:NhLim}
 \end{equation}
 \item For any antenna $i$, the increase in SNR obtained by increasing its number of bits to $b_i'>b_i$ is independent of the state of the other antennas.
 \item If any pair of antennas $i,j$ satisfies $|u_m^i|>|u_m^j|$ but $b_i<b_j$, then the system achieves higher effective SNR and the same power consumption if we swap $b_i$ and $b_j$.
\end{itemize}

The above ideas inspire the Greedy Bit Allocation (GBA) algorithm that allocates bits to ADCs in descending order of the values of $|u_m^i|$, implicitly obtained from the order of $\gamma_i$.

\begin{algorithm}
\small
\caption{Greedy Bit Allocation}
\label{alg:gba}
\begin{algorithmic}
\STATE { Reference effective SNR $\gamma^{\mathrm{ref}}$}
\STATE { Measure unquantized SNR on each antenna $\gamma^i$}
\STATE { Order the RF chains as $\gamma^1\geq\gamma^2\geq\dots \gamma^{N_r}$}
\STATE { Start assuming $N_{\mathrm{high}}=0$, $b_{i}=b_{\mathrm{low}}\forall i$ }
\WHILE { $\sum_{i=1}^{N_r} \gamma_q^i< \gamma_q^{\mathrm{ref}}$ }
  \STATE {$N_{\mathrm{high}}=N_{\mathrm{high}}+1$}
  \STATE {$b_{i}=b_{\mathrm{high}}$}
\ENDWHILE
\end{algorithmic}
\end{algorithm}

The GBA algorithm only allows antennas to use $b_{\mathrm{low}}$ or $b_{\mathrm{high}}$ bits, starts with all ADCs in the low assignment, and swaps to a higher number of bits one antenna at a time until the system has the same effective SNR of the reference. Two outcomes are possible: if at the end of the algorithm the number of high-resolution RF chains is $N_{\mathrm{high}}<\frac{2^{b_{\mathrm{ref}}-b_{\mathrm{low}}}-1}{2^{b_{\mathrm{high}}-b_{\mathrm{low}}}-1}N_r$, power has been saved by GBA. Otherwise, GBA wastes more power than a reference scheme with fixed resolution $b_{\mathrm{ref}}$.

Two weaknesses of GBA are that it does not work for $b_{\mathrm{low}}=b_{\mathrm{ref}}$ and that it always requires to have all antennas active. We add these functionalities to the improved Greedy Antenna Selection and Bit Allocation (GASBA) algorithm.

In the GASBA algorithm, three values are allowed per ADC, 0, $b_{\mathrm{low}}$ or $b_{\mathrm{high}}$ bits. The algorithm combines the mechanics of antenna selection and bit allocation, enabling the receiver to exploit a combination of high-resolution RF chains, lower-resolution RF chains, and disabled RF chains. This algorithm can replicate all allocations possible with GBA and also perform some new allocations in the style of antenna selection.

\begin{algorithm}
\small
\caption{Greedy Antenna Selection and Bit Allocation}
\label{alg:gasba}
\begin{algorithmic}
\STATE { Reference effective SNR $\gamma^{\mathrm{ref}}$}
\STATE { Measure unquantized SNR on each antenna $\gamma^i$}
\STATE { Order the RF chains as $\gamma^1\geq\gamma^2\geq\dots \gamma^{N_r}$}
\STATE { Start assuming $N_{\mathrm{on}}=0$, $N_{\mathrm{high}}=0$, $b_{i}=0\forall i$ }
\WHILE { $\sum_{i=1}^{N_{\mathrm{on}}} \gamma_q^i< \gamma_{\mathrm{ref}}$ }
  \STATE{$N_{\mathrm{on}}=N_{\mathrm{on}}+1$}
  \IF {$N_{\mathrm{high}}<\frac{2^{b_{\mathrm{ref}}-b_{\mathrm{low}}}-1}{2^{b_{\mathrm{high}}-b_{\mathrm{low}}}-1}N_r$}
    \STATE {$N_{\mathrm{high}}=N_{\mathrm{high}}+1$}
    \STATE {$b_{i}=b_{\mathrm{high}}$}
  \ELSE
    \STATE {$b_{i}=b_{\mathrm{low}}$}
  \ENDIF
\ENDWHILE
\end{algorithmic}
\end{algorithm}

\subsection{Initial SNR Measurement}
\label{ssec:SNRMeas}

We assume that channel state information of each receive antenna is perfectly known. In practice, an estimation of the SNR will be needed. The impact of imperfect initial SNR estimation is left as part of our future work. For example in \cite{VeeraVarResADC}, an ADC design is proposed where inputs with lower voltage use 6 bits and inputs with higher voltage use 4 bits. Such a design could be easily modified to operate in the opposite way, giving 6 bits to the signals with the higher voltage. Since the average thermal noise per antenna is the same, the SNR is directly proportional to the squared voltage in the antenna circuits in receivers without Automatic Gain Control (AGC), and is proportional to the ratio between the squared voltage and the amplifier gain in receivers with independent AGC for each antenna.

\section{Numerical Simulations}
In this section we present the power saving achieved by GBA and GASBA algorithms, studied by Monte Carlo simulation with results averaged over 1000 independent realizations. We discuss the performance for different combinations of $b_{\mathrm{low}}, b_{\mathrm{high}}$ for both algorithms. In the simulations, the transmitter is always equipped with 4 antennas whereas the number of receive antennas can be either 64 or 256.
We consider a mmWave link with $B=1$ GHz and vary the unquantized link SNR from $-$20 to 20 dB (except for Figures \ref{fig:GBAN64} and \ref{fig:GBAN256}, where it is varied from $-$20 to 30 dB) with a step of 5 dB. The unquantized SNR is the product of transmitter power and pathloss, divided by the noise power $N_o$.

We display the normalized power consumption $\xi$ \eqref{eq:normpow} vs the unquantized SNR for systems that achieve the same quantized SNR (and thus, capacity). We use references that have 5 or 4 bits resolution, which achieve the best EE of a fixed-resolution system according to \cite{WaqasEConJour16}, \cite{Ahmed16_EE}.

\subsection{GBA}

\begin{figure}[!t]
 \centering
 \subfigure[$N_r = 64$]{
 \includegraphics[width=.70\columnwidth]{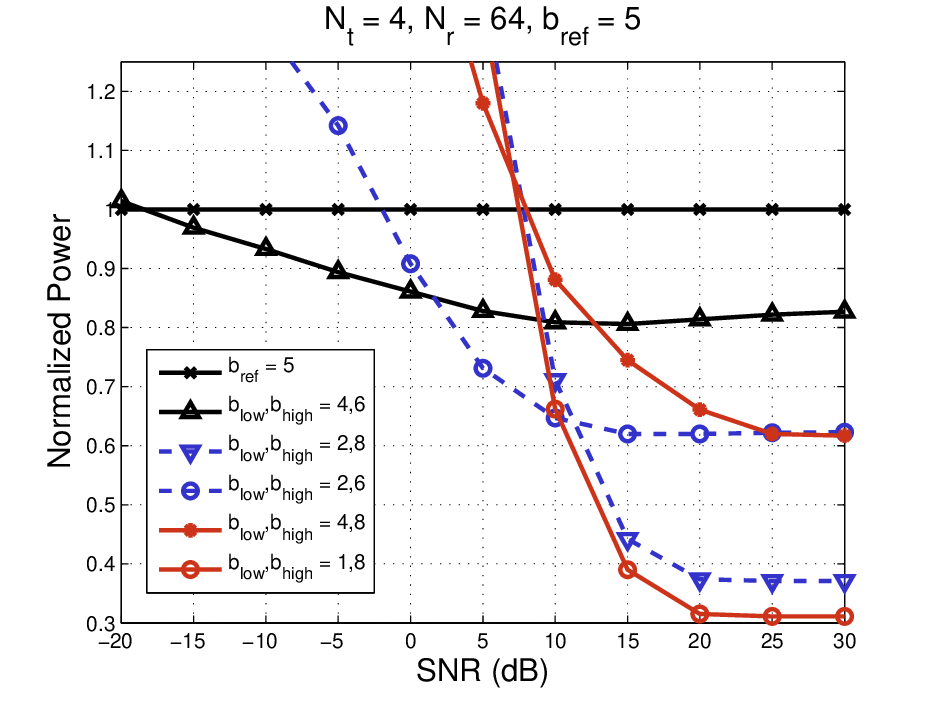}
 \label{fig:GBAN64}
 }
 \subfigure[$N_r = 256$]{
 \includegraphics[width=.70\columnwidth]{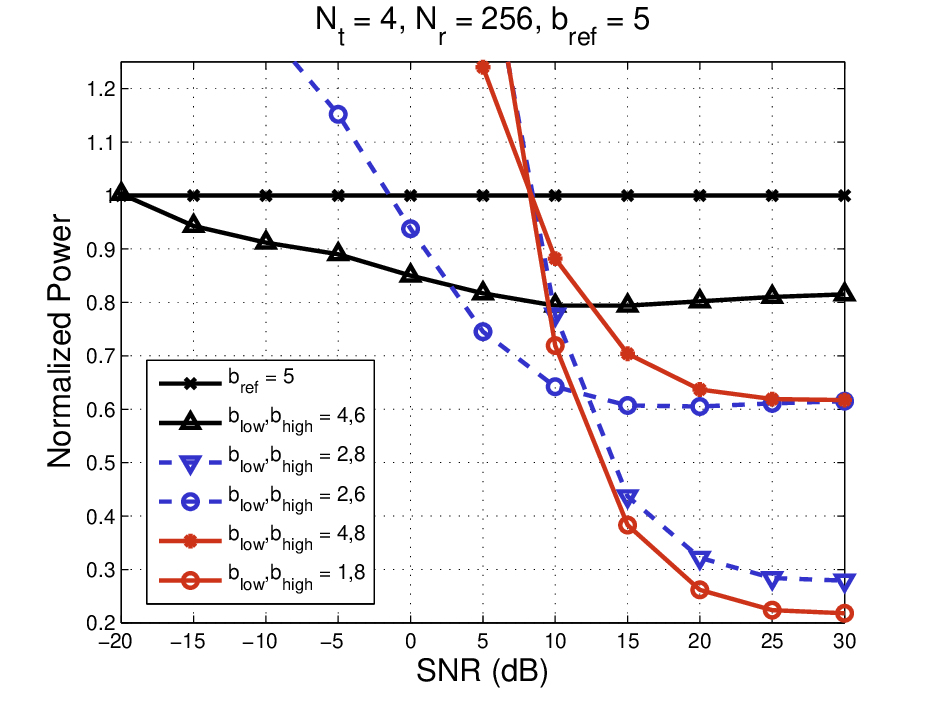}
 \label{fig:GBAN256}
 }
 \caption{Greedy Bit Allocation with $b_{\mathrm{ref}} = 5$}
 \label{fig:GBA}\vspace{-3mm}
\end{figure}

We begin with the results for the GBA algorithm. Figs. \ref{fig:GBAN64} and \ref{fig:GBAN256} show the normalized power consumed by the receivers with GBA for $N_r = 64$ and $N_r = 256$ receive antennas, respectively, and $b_{\mathrm{ref}} =5$.
The result shows that above certain SNR values the variable resolution architecture displays lower power consumption than the fixed resolution architecture.

Note that the configurations where $b_{\mathrm{low}},b_{\mathrm{high}}$ are relatively close to $b_{\mathrm{ref}}$ (e.g., $(b_{\mathrm{low}},b_{\mathrm{high}}) = (4,6)$) result in a lower power consumption for almost the complete range of unquantized SNR.
This is due to the fact that, in circumstances where the quantization noise of the reference is smaller or comparable to the unquantized signal noise, using a high $b_{\mathrm{low}}$ (i.e., close to $b_{\mathrm{ref}}$) already achieves a capacity close to the reference, and thus it only takes very few RF chains with $b_{\mathrm{high}}$ bits to close the gap. 


Secondly, note that in the settings with very low $b_{\mathrm{low}}$ (i.e., with $b_{\mathrm{low}} = 1,2$ bits) a large power is saved at very high SNR, but the variable resolution system consumes even more power than the reference at low SNRs. 
This is due to the fact that if the number of bits is sufficient, a quantized system operates very close to the capacity of an unquantized system, but the threshold that marks this ``sufficient'' number of bits grows with the unquantized SNR \cite{MyPCCompEW16}. Therefore, at high operating SNR the contribution of RF chains with $b_{\mathrm{low}}$ is not significant, and the use of smaller $b_{\mathrm{low}}$ saves power without harming the capacity much, while on the other hand the few ADCs with very high resolution (for instance $b_{\mathrm{high}} = 8$) improve the SNR much more than many RF chains with moderate resolution. Therefore the combination of smaller $b_{\mathrm{low}}$ and higher $b_{\mathrm{high}}$ works better at high SNR.

Conversely, at low SNR, to achieve the capacity of the reference resolution system, a dramatic increase in the number of high resolution ADCs is required, which increases the power consumption of the variables bits scheme even more than the fixed bits reference. For instance, with $(b_{\mathrm{low}},b_{\mathrm{high}})$ set to $(1,8)$, $(2,8)$ and $(4,8)$, the consumed power is lower than the reference only when the SNR is above 10 dB. This is because $b_{\mathrm{high}}$ produces no significant improvement in the SNR and choosing $b_{\mathrm{high}}$ closer to $b_{\mathrm{ref}}$ results in a lower power consumption.

The impact of the number of receive antennas can be observed by comparing Figs. \ref{fig:GBAN64} and \ref{fig:GBAN256}. For instance, with $b_{\mathrm{low}}, b_{\mathrm{high}} = 4,6$, the power of the variable resolution system compared to the fixed resolution reference is in the range $80$-$95$\% for both antenna configurations $N_r=64$ and $N_r = 256$. On the other hand, at the same high SNR of 20 dB, the variable architecture with ($b_{\mathrm{low}}, b_{\mathrm{high}}) = (1,8)$ can achieve a power saving of $69$\% with 64 antennas, whereas the system with 256 antennas can achieve a power saving of $75$\%. It seems that the number of antennas affects the gains more for more extremely separated resolution values that work well at high SNR, whereas it has a negligible impact for narrowly separated resolution values that work well at all SNRs.


Finally, note that there is a certain operating SNR after which the normalized power saturates. 
This is due to the fact that at very high SNR, few high bits ADCs are enough to get the same capacity as a receiver where all antennas are connected to fewer bits ADCs (i.e., with $b_{\mathrm{ref}} = 5$).
Moreover, with ($b_{\mathrm{low}}, b_{\mathrm{high}}) = (4,6)$, the minimum power is achieved at 10 dB and 15 dB SNR for $N_r = 64$ and $N_r = 256$, respectively.  
This is due to the phenomenon explained above that at very low (high) unquantized SNR the contribution of $b_{\mathrm{high}}$ ($b_{\mathrm{low}}$) to the quantized SNR is not very significant.


\begin{figure}[!t]
 \centering
 \subfigure[$b_{\mathrm{ref}} = 5$, $N_r = 64$]{
 \includegraphics[width=.70\columnwidth]{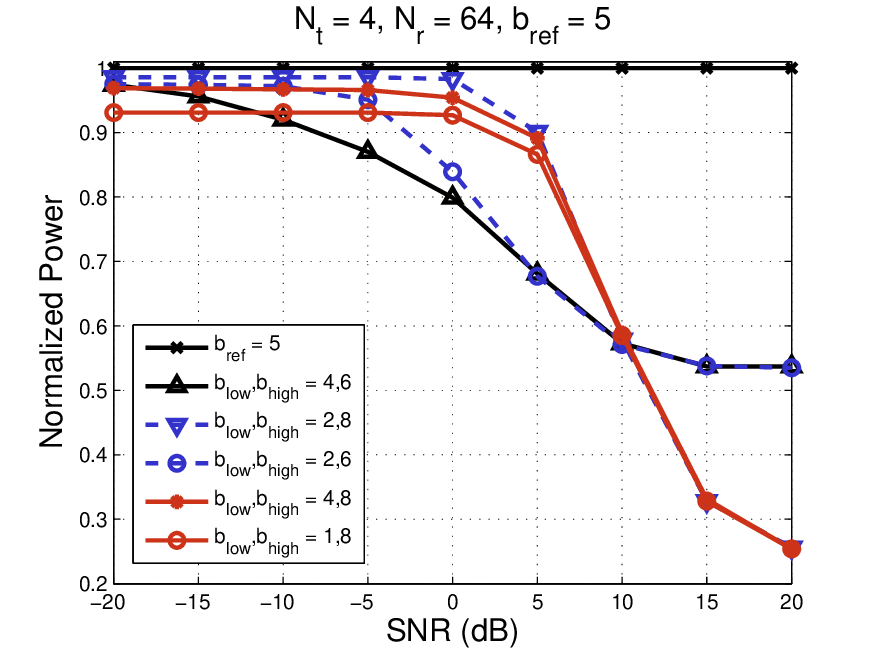}
 \label{fig:GASBAN64Bref5}
 }
 \subfigure[$b_{\mathrm{ref}} = 5$, $N_r = 256$]{
 \includegraphics[width=.70\columnwidth]{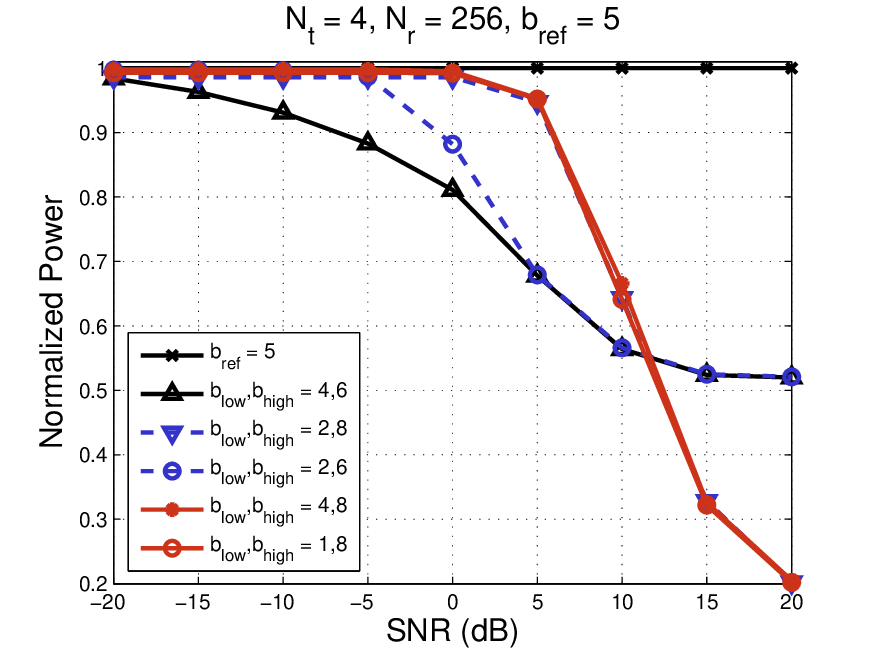}
 \label{fig:GASBAN256Bref5}
 }
 \subfigure[$b_{\mathrm{ref}} = 4$, $N_r = 256$]{
 \includegraphics[width=.70\columnwidth]{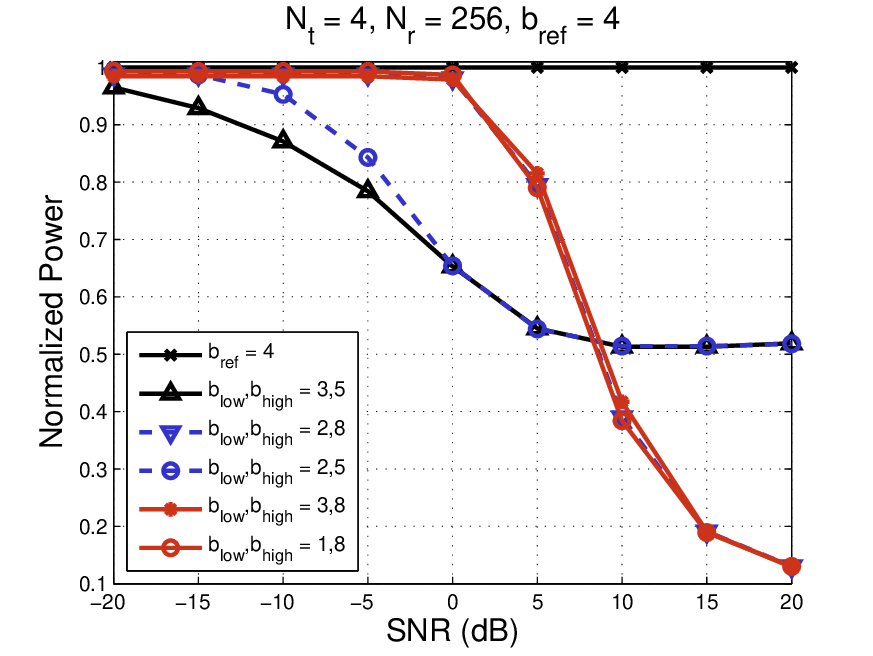}
 \label{fig:GASBAN256Bref4}
 }
 \caption{Greedy Antenna Selection and Bit Allocation}
 \label{fig:GASBA}\vspace{-3mm}
\end{figure}

\subsection{GASBA}

To address the increased power consumption issues of GBA at low SNR, we now discuss the performance of GASBA which also puts constraints on the total power consumed by the high-resolution antennas. 
Note that GBA always tries to achieve the rate of the reference, and sometimes this causes the power consumption to be greater than the reference. However, in GASBA, the maximum $N_{\mathrm{high}}$ (\ref{eq:NhLim}) ensures that the power consumption of the variable architecture is bounded by the reference. Therefore, after reaching the maximum power limit, the GASBA algorithm stops assigning more higher resolution antennas to match the capacity of the reference fixed bit architecture which in some cases may result in a reduced capacity.  

\begin{figure}[!t]
 \centering
 \includegraphics[width=.70\columnwidth]{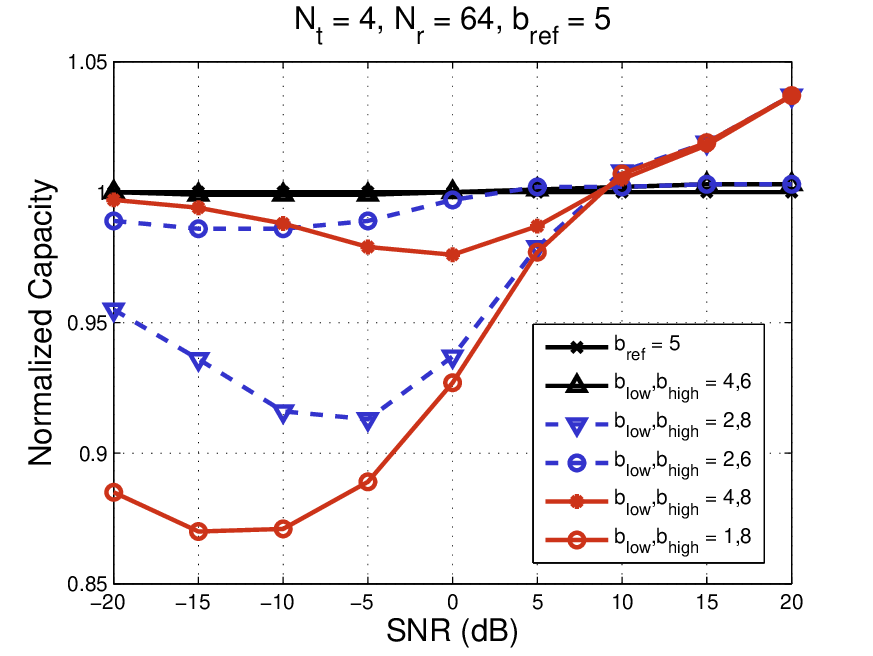}
 \caption{GASBA does not always match the reference capacity at low SNR.}
 \label{fig:GASBAcap}\vspace{-3mm}
\end{figure}

Figs. \ref{fig:GASBAN64Bref5}, \ref{fig:GASBAN256Bref5} show the results for GASBA with $b_{\mathrm{ref}} = 5$, $N_r = 64$ and $N_r = 256$, respectively.
Note that the normalized power of any variable resolution configuration at any operating SNR is less than or equal to the power consumed by the reference architecture, thanks to the design of the algorithm that stops activating high-resolution RF chains when a certain limit is reached. 
At the lowest SNR values, the normalized power consumption of GASBA becomes flat as all antennas are assigned to ADCs, where the number of antennas assigned to $b_{\mathrm{high}}$ ADCs is equal to the maximum value of $N_{\mathrm{high}}$ (i.e., the value of $N_{\mathrm{high}}$ when Eq. \eqref{eq:NhLim} solves with equality) while the rest of the antennas are assigned to $b_{\mathrm{low}}$ ADCs. Note that in these scenarios the GASBA algorithm gives up trying to make the variable-resolution system capacity match the reference, as illustrated in Fig. \ref{fig:GASBAcap}.
Also note that, for some configurations of $b_{\mathrm{low}},b_{\mathrm{high}}$, the value of $N_{\mathrm{high}}$ obtained by solving Eq. \eqref{eq:NhLim} with equality may not be an integer, and therefore in those cases the floor of $N_{\mathrm{high}}$ is selected. Due to this reason the curves for some configurations of $b_{\mathrm{low}},b_{\mathrm{high}}$ stay below the reference even when all antennas are utilized (Figure \ref{fig:GASBA}).

The power savings improve at high SNR. This is due both to the effects described for GBA above, and to the fact that a system with higher captured energy per antenna does not need so much beamforming gain and can make do with activating fewer RF chains. In fact, at higher SNRs using only the RF chains with $b_{\mathrm{high}}$ bit ADCs is enough to provide equal capacity as the reference, and the GASBA algorithm greatly reduces the subset of selected antennas at high SNR, keeping power consumption minimal.

The results show that at low SNRs the configurations with $b_{\mathrm{low}}$ and $b_{\mathrm{high}}$ closer to $b_{\mathrm{ref}}$ (i.e., $(b_{\mathrm{low}}, b_{\mathrm{high}}) = (4,6)$ or $(2,6)$) are the only ones that can always achieve the same capacity as the reference. 
Moreover, these configurations also result in some power consumption savings at low SNR (e.g., at SNR = 0 dB). 
This follows from the fact that at low SNR too many bits do not provide any significant improvement, which instead could be obtained by increasing the number of antennas. A smaller $b_{\mathrm{high}}$ results in a higher upper limit to $N_{\mathrm{high}}$, and therefore the use of many moderately-high resolution antennas is preferable over few very high resolution ones. 

Conversely, as the SNR increases it is preferable to use few antennas with many bits. At higher SNRs, the systems with very high $b_{\mathrm{high}}$ start to display the best power savings. However, these values must not be too extreme for a given SNR. For example, at 10 dB the configuration $(2,6)$ actually performs better than $(2,8)$. Nonetheless, the curves for $b_{\mathrm{high}} = 8$ display a more abrupt decrease. The value of  $b_{\mathrm{low}}$ is irrelevant at high SNR because those RF chains are never turned on, differently from GBA where it had some small impact.

Note that in the entire range $-$20 to 20 dB SNR, which is quite reasonable for wireless communication, the variable resolution scheme with $(b_{\mathrm{low}}, b_{\mathrm{high}}) = (4,6)$ always results in a reasonable power reduction without any degradation in the achievable capacity for both GASBA and GBA algorithms. Thus, the variable-resolution model in\cite{MixedADC} with $b_{\mathrm{low}}=1$, despite showing impressive gains at very high SNR, may not be the best configurations for all systems.

Note that although the GASBA algorithm always results in a lower power consumption than the reference fixed resolution  architecture, there are regimes where GBA consumes lower power than GASBA for the same number of bits configuration in their ADCs.
For instance, with 256 antennas the configuration $(b_{\mathrm{low}}, b_{\mathrm{high}}) = (4,6)$ at $-$10 dB SNR displays a normalized power of 0.90 (Fig. \ref{fig:GBAN256}) and 0.95 (Fig. \ref{fig:GASBAN256Bref5}) for GBA and GASBA, respectively.
This is because the GASBA algorithm starts by trying to approximate the capacity of the reference using only $b_{\mathrm{high}}$ ADCs and therefore in certain regimes GASBA may assign high-resolution ADCs to antennas where low-resolution ADCs could have done the same job (low-SNR antennas where the quantization noise is small or negligible compared to the thermal noise).

Finally, in Fig. \ref{fig:GASBAN256Bref4} we observe the effect of changing the reference for a variable resolution GASBA system compared versus $b_{\mathrm{ref}} = 4$.
The results show that with a reduction in $b_{\mathrm{ref}}$ the difference in power consumption between the fixed and variable resolution techniques increases (compare Figs. \ref{fig:GASBAN256Bref5} and \ref{fig:GASBAN256Bref4}). This is because a reduction in $b_{\mathrm{ref}}$ also reduces the target reference capacity  and therefore the required number of ADCs with $b_{\mathrm{high}}$ bits decreases.

In summary, the use of variable resolution ADCs can provide a significant decrease in the power consumption in comparison to a fixed resolution ADC architecture, ranging from 20\% at 0 dB SNR up to 80\% at 20 dB.

\section{Conclusions}

In this paper, we proposed variable resolution quantization in fully digital receiver architectures with large antenna arrays. This variable-resolution ADC approach can be seen as a generalization of antenna selection and other 1-bit ADC proposals.

We discussed models for a mmWave uplink scenario with a scalar transmitted signal and beamforming over a sparse scattering channel matrix where a single dominant eigenvalue is responsible for most of the energy transfer of the system. We have noted the usual power-consumption model for ADCs, and proposed a simple ``first-attempt'' type of two-level resolution ADC system design. We have also designed two algorithms to operate in our model, one that merely alternates between high and low resolution states for the ADCs, and one that adds a third off state inspired by antenna-selection techniques.

We have studied the capacity and power consumption of the mmWave link under this variable resolution model, and have shown that there can be very significant power savings up to 80\% in variable-resolution quantization schemes, depending on the link SNR pre-quantization. Our results also show that the benefits are greater with not-so-low and not-so-high numbers of bits for the low and high resolution levels of the ADCs, respectively. This approach can outperform the existing literature proposing variable resolution systems with 1 bit for the lowest resolution ADCs.

We would also like to point out that, even though the power savings obtained in our results may not seem so impressive in the lower SNR regimes, such low-SNR systems do not usually benefit from the kind of spatial multiplexing gains that are the staple of Digital Combining. Therefore, instead of choosing variable-vs-fixed resolution ADCs, at lower SNRs a more radical switch to Analog Combining would make more practical sense.

In the future, we will extend the analysis of variable resolution ADCs design with full spatial multiplexing and optimize the choice of the variable bits based on the power and capacity contraints.   

\bibliographystyle{IEEEtran}
\bibliography{mmWave,otherrefs}

%
%
%
%

\end{document}